\documentclass[12pt]{article}
\usepackage{graphicx}

\def\Re{{\cal R \mskip-4mu \lower.1ex \hbox{\it e}\,}}
\def\Im{{\cal I \mskip-5mu \lower.1ex \hbox{\it m}\,}}
\def\ie{{\it i.e.}}
\def\eg{{\it e.g.}}
\def\etc{{\it etc}}
\def\etal{{\it et al.}}

\def\sub#1{_{\lower.25ex\hbox{$\scriptstyle#1$}}}
\def\tev{\,{\ifmmode\mathrm {TeV}\else TeV\fi}}
\def\gev{\,{\ifmmode\mathrm {GeV}\else GeV\fi}}
\def\mev{\,{\ifmmode\mathrm {MeV}\else MeV\fi}}
\def\mpl{\ifmmode M_{pl}\else $M_{pl}$\fi}
\def\to{\rightarrow}

\def\subw{_{\rm w}}
\def\mh{\ifmmode m\sbl H \else $m\sbl H$\fi}
\def\mch{\ifmmode m_{H^\pm} \else $m_{H^\pm}$\fi}
\def\mt{\ifmmode m_t\else $m_t$\fi}
\def\mc{\ifmmode m_c\else $m_c$\fi}
\def\mz{\ifmmode M_Z\else $M_Z$\fi}
\def\mw{\ifmmode M_W\else $M_W$\fi}
\def\mws{\ifmmode M_W^2 \else $M_W^2$\fi}
\def\mhs{\ifmmode m_H^2 \else $m_H^2$\fi}   
\def\mzs{\ifmmode M_Z^2 \else $M_Z^2$\fi}
\def\mts{\ifmmode m_t^2 \else $m_t^2$\fi}
\def\mcs{\ifmmode m_c^2 \else $m_c^2$\fi}
\def\mchs{\ifmmode m_{H^\pm}^2 \else $m_{H^\pm}^2$\fi}
\def\ztwo{\ifmmode Z_2\else $Z_2$\fi}
\def\zone{\ifmmode Z_1\else $Z_1$\fi}
\def\mtwo{\ifmmode M_2\else $M_2$\fi}
\def\mone{\ifmmode M_1\else $M_1$\fi}
\def\tb{\ifmmode \tan\beta \else $\tan\beta$\fi}
\def\xw{\ifmmode x\subw\else $x\subw$\fi}
\def\ch{\ifmmode H^\pm \else $H^\pm$\fi}
\def\lum{\ifmmode {\cal L}\else ${\cal L}$\fi}
\def\inpb{\,{\ifmmode {\mathrm {pb}}^{-1}\else ${\mathrm {pb}}^{-1}$\fi}}
\def\infb{\,{\ifmmode {\mathrm {fb}}^{-1}\else ${\mathrm {fb}}^{-1}$\fi}}
\def\epem{\ifmmode e^+e^-\else $e^+e^-$\fi}
\def\ppb{\ifmmode \bar pp\else $\bar pp$\fi}
\def\bsg{\ifmmode B\to X_s\gamma\else $B\to X_s\gamma$\fi}
\def\bsll{\ifmmode B\to X_s\ell^+\ell^-\else $B\to X_s\ell^+\ell^-$\fi}
\def\bstt{\ifmmode B\to X_s\tau^+\tau^-\else $B\to X_s\tau^+\tau^-$\fi}
\def\lamt{\ifmmode \tilde\lambda\else $\tilde\lambda$\fi}
\def\shat{\ifmmode \hat s\else $\hat s$\fi}
\def\that{\ifmmode \hat t\else $\hat t$\fi}
\def\uhat{\ifmmode \hat u\else $\hat u$\fi}

\newskip\zatskip \zatskip=0pt plus0pt minus0pt
\def\matth{\mathsurround=0pt}

\def\atversim#1#2{\lower0.7ex\vbox{\baselineskip\zatskip\lineskip\zatskip
  \lineskiplimit 0pt\ialign{$\matth#1\hfil##\hfil$\crcr#2\crcr\sim\crcr}}}

\renewcommand{\thefootnote}{\fnsymbol{footnote}}

\begin{document} \begin{titlepage} 
\rightline{\vbox{\halign{&#\hfil\cr
&SLAC-PUB-9295\cr
&August 2002\cr}}}
\begin{center}

{\Large\bf Unique Identification of Graviton Exchange Effects in $e^+e^-$ 
Collisions}
\footnote{Work supported by the Department of 
Energy, Contract DE-AC03-76SF00515}
\medskip

\normalsize 
{\bf \large Thomas G. Rizzo}
\vskip .3cm
Stanford Linear Accelerator Center \\
Stanford University \\
Stanford CA 94309, USA\\
\vskip .2cm

\end{center}

\begin{abstract}
Many types of new physics can lead to contact interaction-like modifications 
in $e^+e^-$ processes below direct production threshold. We examine the 
possibility of uniquely identifying the effects of graviton exchange, which are 
anticipated in many extra dimensional theories, from amongst this large set of 
models by using the moments of the angular distribution of the final state 
particles. In the case of the $e^+e^-\to f\bar f$ process we demonstrate that 
this technique allows for the unique identification of the graviton exchange 
signature at the $5\sigma$ level for mass scales as high as $6\sqrt s$. The 
extension of this method to the $e^+e^-\to W^+W^-$ process is also discussed.
\end{abstract} 



\renewcommand{\thefootnote}{\arabic{footnote}} \end{titlepage}


\section{Introduction}

It is generally expected that new physics beyond the Standard Model(SM) will 
manifest itself at future colliders that probe the TeV scale such as the LHC 
and the Linear Collider. This new physics(NP) may appear either directly, 
as in the case of new particle production, \eg, SUSY, or indirectly through 
deviations from the predictions of the SM. In the case of direct production, 
the discovery and 
identification of the NP would be relatively straightforward once masses and 
various couplings were determined through precision measurements. 
In the case of indirect discovery the 
effects may be subtle and many different NP scenarios may lead to the same or 
similar experimental signatures. Clearly, identifying the origin of the NP in 
these circumstances will prove more difficult and new tools must be available 
to deal with this potentiality. 

Perhaps the most well known example of this indirect scenario in a 
collider context would be the observation of deviations 
in, \eg, various $e^+e^-$ cross sections due to apparent contact 
interactions{\cite {elp}}. There are many very different NP scenarios 
that predict new particle exchanges which can lead to contact interactions 
below direct production threshold; 
a partial list of known candidates is: a $Z'$ from an extended 
gauge model{\cite {e6,zp}}, scalar or vector  
leptoquarks{\cite {e6,lq}, $R$-parity violating sneutrino($\tilde \nu$) 
exchange{\cite {rp}}, scalar or vector 
bileptons{\cite {bl}}, graviton Kaluza-Klein(KK) 
towers{\cite {ed,dhr}} in extra dimensional models{\cite {add,rs}}, 
gauge boson KK towers{\cite {ed2,dhr}}, and even string 
excitations{\cite {se}}. Of course, there may be many other 
sources of contact interactions from NP models as yet undiscovered, as was 
low-scale gravity only a few years ago. 

If contact interaction effects are observed one can always try to fit the 
shifts in the observables to each one of the set of known theories and see 
which gives the best fit--an intensive approach followed by Pasztor and 
Perelstein{\cite {gm}}. Identifying the model that fits best, it may then be 
possible to select a starting point for further exploration and model building. 
Alternatively, it may be useful to devise a test or set of 
tests which will rather quickly divide the full set of all possible 
models into subclasses 
which can then be studied further by other techniques. In particular, it would 
be useful to have a method that rapidly identifies basic features about certain 
model classes. In this paper we propose such a technique that makes use of the 
specific modifications in angular distributions induced by $s-$ and $t-$channel 
exchanges of particles of various spins. As we will see below this method 
offers a way to uniquely identify graviton KK tower exchange 
(or, indeed, $\tilde \nu$ or any other possible spin-0 exchange)
provided it is dominant the source of the new contact interaction.

\section{Technique}

In order to introduce our technique, let us consider the normalized cross 
section for the process 
$e^+e^- \to f\bar f$ in the SM assuming $m_f=0$ and $f\neq e$ 
for simplicity. This can be written as 
\begin{equation}
{{1}\over {\sigma}} {{d\sigma}\over {dz}}={3\over {8}}(1+z^2)+A_{FB}(s)z\,,
\end{equation}
where $z=\cos \theta$ and $A_{FB}(s)$ is the Forward-Backward Asymmetry which 
depends upon the electroweak quantum numbers of the fermion, $f$, as well as 
the center of mass energy of the collision, $\sqrt s$. This structure is 
particularly interesting in that it is equally valid for a wide variety of 
New Physics models: composite-like contact interactions, $Z'$ models, 
TeV-scale KK gauge bosons, as well 
as for $t-$ or $u-$ channel leptoquark and bilepton exchanges at 
leading order in $s/M^2$, where $M$ is the leptoquark or bilepton mass. 
In fact, for this observable the {\it only} deviation from the 
SM for any of these models will be through the variations in  
the value of $A_{FB}(s)$ since we have chosen to normalize the cross section. 

Now let us consider taking moments of the normalized cross section above with 
respect to the Legendre Polynomials, $P_n(z)$. This can be done easily by 
re-writing Eq.(1) as 
\begin{equation}
{{1}\over {\sigma}} {{d\sigma}\over {dz}}={1\over {2}}P_0+{1\over {4}}P_2
+A_{FB}(s)P_1\,,
\end{equation}
and recalling that the $P_n(z)$ are normalized as 
\begin{equation}
\int_{-1}^{1} ~dz P_n(z) P_m(z)={2\over {2n+1}}\delta_{nm}\,. 
\end{equation}
Denoting such moments as $<P_n>$, one finds that $<P_1>=2A_{FB}/3$, 
$<P_2>=1/10$ and $<P_{n>2}>=0$. In addition we also trivially obtain that 
$<P_0>=1$ since we have normalized the distribution so that this moment 
carries no new information. Thus, very naively, if 
we find that the $<P_{n>1}>$ are given by their 
SM values while $<P_1>$ differs 
from its corresponding SM value we could conclude that the NP is most likely 
one of those listed immediately 
above. If {\it both} $<P_{1,2}>$ differ from their SM expectations 
while the $<P_{n>2}>$ remain zero the source can only be $\tilde \nu$, or more 
generally, a 
spin-0 exchange in the $s-$channel. As we will see below only $s-$channel 
KK graviton exchange, since it is spin-2, leads to non-zero values of 
$<P_{3,4}>$ while the $<P_{n>4}>$ still remain zero. Of course the values of 
$<P_{1,2}>$ will also be different from their SM values in this case but 
as we have just observed this signal is not unique to gravity. 
This observation seems to yield a rather simple test for the 
exchange of graviton KK towers{\cite {missing}}. It is important to note that 
we could not have performed this simple analysis 
for the case of Bhabha scattering, \ie, $e^+e^- 
\to e^+e^-$, as it involves both $s-$ and $t-$channel exchanges in the SM and 
thus {\it all} of the $<P_n>$ would be non-zero. 

Of course, the real world is not so simple as the idealized case 
we have just discussed for 
several reasons. First, we have assumed that we know the cross section 
precisely at all values of $z$, \ie, we have infinite statistics with no 
angular binning. 
Secondly, to use the orthonormality conditions above we need to have complete 
angular coverage, \ie, no holes for the beam pipe, \etc. 
To get a feeling for how 
important these effects can be let us first consider dividing the distribution 
into a finite number of angular bins, $N_{bins}$, of common size 
$\Delta z=2/N_{bins}$. Instead of 
doing a simple integral we must perform a sum, \ie, we make the replacement  
\begin{equation}
\int_{-1}^1 dz P_n(z){{1}\over {\sigma}} {{d\sigma}\over {dz}}\to \sum_{bins} 
P_n(z_i) \sigma_i/\sigma\,,
\end{equation}
where $i$ labels the bin number, 
$\sigma_i$ is the cross section in each bin obtained by direct integration 
and $z_i$ is the bin center at which the $P_n$ are to be evaluated. The 
results of this analysis are shown in Table 1 for the case of large  
statistics; here we see that as the number 
of bins grows large we rapidly recover the continuum results discussed above. 
Of course in any realistic experimental situation, $N_{bins}$ remains finite
but we see that a value of order 20 is reasonable as it strikes a respectable 
balance between the realistic 
demands of statistics, angular resolution and taking $N_{bins}$ sufficiently 
large. The fact that we do not recover the trivial SM results 
above in this case can be considered as a `background' in a loose sense. We 
will return to this point below.

\begin{table}
\centering
\begin{tabular}{|c|c|c|c|c|} \hline\hline
$N_{bins}$&$<P_2>(10^{-2})$&$<P_4>(10^{-3})$&$<P_1>$  &$<P_3>(10^{-3})$  
\\ \hline \hline
   10   & 9.0040 &-26.7585    & 0.66000    &-23.1000      \\
   20   & 9.7503 & -6.8285    & 0.66500    & -5.8188      \\
   50   & 9.9600 & -1.0988    & 0.66640    & -0.9330      \\
  200   & 9.9975 & -0.0687    & 0.66665    & -0.0583      \\
 1000   & 9.9999 & -0.0027    & 0.66667    & -0.0023      \\
$\infty$  & 10.0   &    0.0     &  2/3       &     0.0      \\ \hline\hline
\end{tabular}
\caption{Dependence on $N_{bins}$ for the first four moments of the 
normalized cross section appearing on the  
right hand side of Eq.(1). Both $<P_{1,3}>$ are in units of $A_{FB}$.} 
\end{table}

Now let us assume that $N_{bins}=20$ and examine the effects of the necessary 
cut at small angles due to the beam pipe, \etc. (Of course this cut is made 
symmetrically near both $0^o$ and $180^o$ so as to not induce additional 
backgrounds into the moments.) This is straightforward to 
implement from the above and leads to the results 
shown in Table 2 for various values of the small angle cut. We note that by 
including this cut the value of $z_i$ at which the $P_n$ are evaluated changes 
for the two bins nearest the beam pipe at either end of the detector as we 
always assume they are to be evaluated at the center of the relevant range 
in $z$.  Here we observe that the `background contamination' of the naive SM 
result increases quite rapidly as we make the angular cut stronger. 

\begin{table}
\centering
\begin{tabular}{|c|c|c|c|c|} \hline\hline
Cut(mr)&$<P_2>(10^{-2})$&$<P_4>(10^{-3})$&$<P_1>$  &$<P_3>(10^{-3})$  
\\ \hline \hline
   0    & 9.7503 & -6.8285    & 0.66500    & -5.8188      \\
   10   & 9.7428 & -6.8981    & 0.66490    & -5.9156      \\
   50   & 9.5652 & -8.5590    & 0.66251    & -8.2301      \\
  100   & 9.0159 & -13.616    & 0.65508    & -15.341      \\
  200   & 6.9030 & -31.895    & 0.62600    & -41.974      \\ \hline\hline
\end{tabular}
\caption{Dependence on the cut at small scattering angles in milliradions 
assuming $N_{bins}=20$ for the first four moments of the normalized cross 
section appearing on the right 
hand side of Eq.(1). Both $<P_{1,3}>$ are in units of $A_{FB}$.} 
\end{table}

What this brief study indicates is that for a realistic detector at a linear 
collider the simple and naive expectations for the various moments will receive 
`backgrounds' that will need to be dealt with and subtracted from the real 
data to obtain information on the $<P_n>$. In the real world these 
backgrounds can be found for a given detector through a detailed Monte 
Carlo(MC) study whose results will be influenced the detector geometry 
and by how well the properties of the detector are known. For our 
numerical analysis below we will follow a simpler approach by calculating the 
moments in the SM (after binning and cuts are applied) and then subtracting 
them from those obtained when the NP is present. In a more realistic 
analysis this means that we will 
assume that the detailed detector MC study can determine these backgrounds 
with reasonably high precision so that they can be subtracted once the 
actual data is available for analysis.

\section{Analysis}

Given the discussion above it is clear that we should begin by examining the 
process $e^+e^-\to f\bar f$; we will return to other potentially interesting 
processes below. To be specific we will concentrate on the 
model of Arkani-Hamed \etal{\cite {add}}, ADD, though our results are easily 
extended to the case of the the Randall-Sundrum model{\cite {rs}} below the  
graviton resonance production threshold. 
The differential cross section for $e^+e^-\to f\bar f$, now including graviton 
tower exchange, for massless fermions can be written as{\cite {ed}}  
\begin{eqnarray}
\label{dsdz}
{d\sigma\over dz} & = & N_c{\pi\alpha^2\over s}\left\{ \tilde P_{ij}
\left[A^e_{ij}A^f_{ij}(2P_0+P_2)/3+2B^e_{ij}B^f_{ij}P_1\right]\right.
\nonumber\\
& & -{\lambda s^2\over 2\pi\alpha \Lambda_H^4}\tilde P_i
\left[v^e_iv^f_i~(2P_3+3P_1)/5+a^e_ia^f_i P_2 \right] \\
& & \left. +{\lambda^2s^4\over 16\pi^2\alpha^2 \Lambda_H^8}\left[
(16P_4+5P_2+14P_0)/35\right]\right\} \,,
\nonumber
\end{eqnarray}
where the indices $i,j$ are summed over the $\gamma$ and $Z$ exchanges, 
$z=\cos \theta$ as above, $\tilde P_{ij}$ and 
$\tilde P_i$ are the usual dimensionless propagator factors (defined in \eg,
\cite{e6}), $A^f_{ij}=(v_i^fv_j^f
+a_i^fa_j^f)\,, B^f_{ij}=(v_i^fa_j^f+v_j^fa_i^f)\,, P_n=P_n(z)$ and $N_c$ 
represents the number of colors of the final state. $\Lambda_H$ is the cutoff 
scale employed by Hewett{\cite {ed}} in evaluating the summation over the 
tower of KK 
graviton propagators and $\lambda=1$ will be assumed in what follows. Our 
results will not depend upon this particular choice of sign. 
In this expression we 
explicitly see the dependence on the $P_{n>2}$ associated with the exchange 
of the tower of KK gravitons. Note that term proportional to $P_{3}$ occurs 
in the interference between the SM and gravitational contributions whereas the 
term proportional to $P_{4}$ occurs only in the pure gravity piece. This 
implies that for $\sqrt s <<\Lambda_H$ it will be $<P_{3}>$ which 
will show the largest shifts from the expectations of the SM. 
With the polarized beams that we expect to have available at a linear 
collider, a z-dependent Left-Right Asymmetry, $A_{LR}$, can also 
be formed and provides an additional observable; this is proportional to the 
difference $d\sigma_L-d\sigma_R$ with $\sigma_{L,R}$ being the 
cross section obtained with left- or right-handed polarized electrons. Using 
the notation above this asymmetry can be written as 
\begin{eqnarray}
\label{alr}
A_{LR}(z) & = & \tilde P_{ij}\left[ 
B^e_{ij}A^f_{ij}(P_2+2P_0)/3+A^e_{ij}B^f_{ij}P_1
\right]/D\\
& & -{\lambda s^2\over 2\pi\alpha \Lambda_H^4}
\tilde P_i\left[a^e_iv^f_i~(2P_3+3P_1)/5+v^e_ia^f_iP_2 \right]/D\,,\nonumber
\end{eqnarray}
where $D$ is given by the curly bracket in the cross section expression 
above. Note that in the 
presence of graviton exchange this quantity also explicitly depends upon 
the $P_{n>2}$ with the leading corrections again expected in $<P_3>$. 

Our approach will be as follows: we consider two observables ($i$) the 
normalized unpolarized cross section and ($ii$) the normalized difference 
of the polarized cross sections $\sim (d\sigma_L-d\sigma_R)/dz$, which is 
essentially given by the numerator terms in the expression for $A_{LR}$. We 
then calculate the first four non-trivial moments of these two observables 
for the $\mu,\tau, b,c$ and $t$ final states within the SM including the 
effects of Initial State Radiation(ISR). (Note that for the $t\bar t$ 
final state we need to generalize the expressions above to include finite 
mass effects. This means that for $t\bar t$ all of the moments will become 
$\sqrt s$ dependent asymptoting to the values given above as 
$\sqrt s \to \infty$.) Here we will assume tagging 
efficiencies of $100\%$, $100\%$, $80\%$, $60\%$ and 
$60\%$, respectively, for the various final states and that $N_{bins}=20$ with 
$\theta_{cut}=$50mr for purposes of demonstration. The resulting values for 
the $<P_n>$ as calculated in the SM will be called `background' values 
consistent with our discussion above. Next, we calculate the 
same moments in the ADD model by choosing a  
value for the parameter $\Lambda_H$. Combining both observables and summing 
over the various flavor final states we can form a $\chi^2$ from the 
deviation of the $<P_{3,4}>$ moments from their SM `background'  
values. For a fixed integrated 
luminosity this can be done using the statistical errors 
as well as the systematic errors associated with the precision expected on 
the luminosity and polarization measurements. (Here we will assume the values 
$\delta L/L=0.25\%$ and $\delta P/P=0.3\%$ in order to 
incorporate these systematic 
effects.) Next we vary the value of the scale $\Lambda_H$ until we obtain a 
$5\sigma$ deviation from the SM; we call this value of $\Lambda_H$ the 
Identification Reach as it is the maximum value for the scale at which we 
observe a $5\sigma$ deviation from the SM values of $<P_{3,4}>$ which we now 
know can only arise due to the effects of graviton exchange. Note that 
this value of the scale should {\it not} be confused with the Discovery Reach 
at which one observes an overall deviation from the SM. Here we are 
specifically looking 
only at deviations in these special moments since they alone are graviton 
sensitive. Although both the $<P_{1,2}>$ {\it also} deviate from their SM 
values these shifts cannot be directly attributed to a spin-2 exchange. (As 
noted previously, the shift in $<P_{1}>$ results in any of the NP models listed 
above whereas a shift in $<P_{2}>$ occurs whenever the new $s$-channel 
exchange is {\it not} spin-1, \eg, $\tilde \nu$ exchange.)

\begin{table}
\centering
\begin{tabular}{|c|c|c|c|c|} \hline\hline
 $f$  &$<P_2>(10^{-2})$&$<P_4>(10^{-3})$&$<P_1>$  &$<P_3>(10^{-3})$  
\\ \hline \hline
$\mu,\tau$ & 9.58 & -8.58    & 0.319    & -3.97      \\
 $b$       & 9.58 & -8.58    & 0.419    & -5.21      \\
 $c$       & 9.58 & -8.58    & 0.407    & -5.06      \\
 $t$       & 4.41 & -6.87    & 0.269    & -3.34      \\ \hline\hline
\end{tabular}
\caption{SM values for the various moments of the normalized unpolarized 
cross section for various flavors as $\sqrt s=500$ GeV. Only the top quark is 
assumed massive. We take $N_{bins}=20$ and $\theta_{cut}=50$mr.} 
\end{table}

To first get an idea of the influence of the graviton KK exchange consider the 
results shown in Tables 3 and 4. The first of these Tables shows the moment 
values in the SM for several final state flavors at a $\sqrt s=500$ GeV 
collider. Note that the background value of 
$<P_{3}>$ is proportional to $<P_{1}>$ for each flavor due to `leakage' 
and that the moments 
for the top quark differ significantly from the others due to the large 
finite mass effects. The proportionality of $<P_{3}>$ and $<P_{1}>$ for all 
flavors signals that $<P_{3}> \neq 0$ is arising from the background and not 
from NP sources.
In the second Table we see that when the graviton KK contributions 
are turned on there are respectable shifts in $<P_{1,2}>$ for all flavors of 
order $10-30\%$ while the corresponding shifts in $<P_{4}>$ are 
somewhat smaller. On the 
otherhand the deviations in $<P_{3}>$ are at the $100-400\%$ level and 
even changes in sign are observed. As expected, $<P_{3}>$ shows the greatest 
sensitivity to graviton exchange. Note that the values of $<P_{3}>$ and 
$<P_{1}>$ are no longer correlated for the different flavors as is the case 
for the SM background. Clearly the large shifts in $<P_3>$ and the fact that 
the values of $<P_3>/<P_1>$ are now flavor dependent 
and differ significantly from 
their SM values is a unique signature for KK graviton exchange.

\begin{table}
\centering
\begin{tabular}{|c|c|c|c|c|} \hline\hline
 $f$  &$<P_2>(10^{-2})$&$<P_4>(10^{-3})$&$<P_1>$  &$<P_3>(10^{-3})$  
\\ \hline \hline
$\mu,\tau$ & 8.41 & -8.03    & 0.286    & -12.69      \\
 $b$       & 5.41 & -6.65    & 0.376    & -16.04      \\
 $c$       & 11.76 & -9.00    & 0.448    & 5.93      \\
 $t$       & 5.45 & -7.14    & 0.283    & 3.92     \\ \hline\hline
\end{tabular}
\caption{Same as the previous table but now assuming $\Lambda_H=2$ TeV.}
\end{table}

Returning to our calculation outlined above we can straightforwardly determine 
the ID reach; this is shown in Fig.1 for several values of $\sqrt s$ as a 
function of the integrated luminosity. Specifically, 
for a $\sqrt s=500$ GeV machine with 
an integrated luminosity of $1~ab^{-1}$ the ID reach with single(double) beam 
polarization is found to be 2.6(3.0) TeV, \ie, $(5-6)\sqrt s$. We remind the 
reader that the corresponding search reach for these luminosities is in 
range of $(9-10)\sqrt s${\cite {tdr}}. Note that the ID reach obtained by 
this approach is a rather 
respectable fraction, more than half, of the search reach.

\begin{figure}[htbp]
\centerline{
\includegraphics[width=9cm,angle=90]{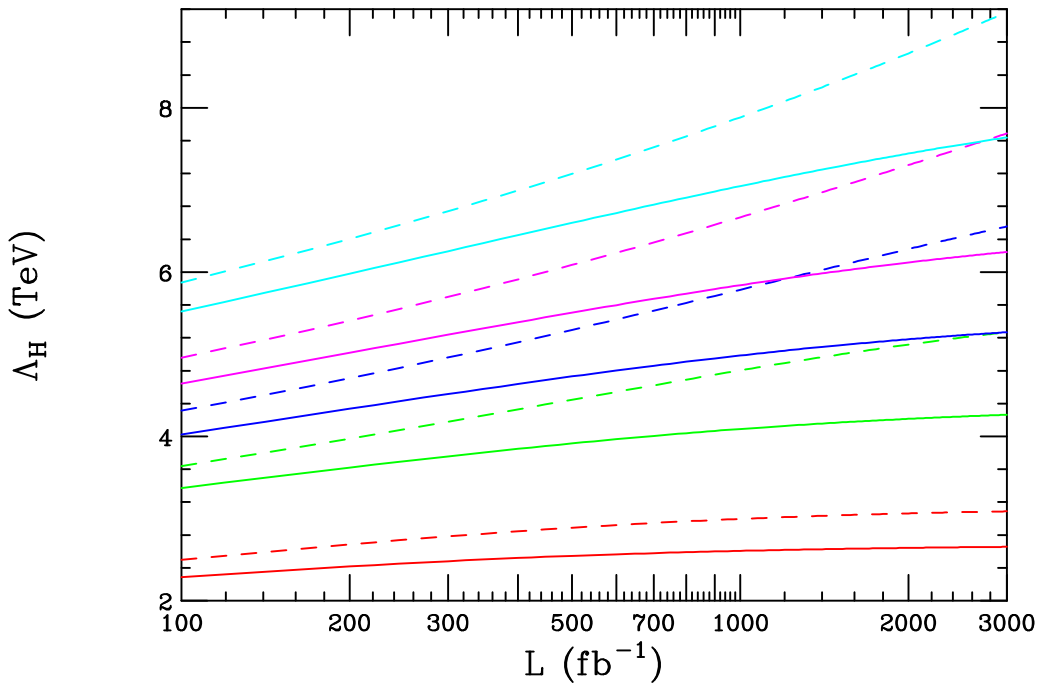}}
\vspace*{0.1cm}
\caption{Identification reach in $\Lambda_H$ as a function of the integrated 
luminosity fro the process $e^+e^-\to f\bar f$, with $f$ summed over 
$\mu,\tau,b,c$ and $t$. The solid(dashed) curves are for an $e^-$ polarization 
of $80\%$(together with a $e^+$ polarization of $60\%$). From bottom to top 
the pairs of curves are for $\sqrt s=0.5, 0.8,1, 1.2$ and $1.5$ TeV, 
respectively.}
\end{figure}

Before turning to other possible processes, it is instructive to examine what 
influence other NP scenarios might 
have on the $<P_n>$. An artificial scenario that 
would most closely mimic gravity would be the $s-$channel exchange of a 
universally coupled scalar field. Like gravity, since the scalar exchange does 
not interfere with the $\gamma$ and $Z$ SM contributions, the leading effects 
are effectively 
from dimension-8 operators. To be specific, we assume $\sqrt s=500$ GeV, 
a Yukawa coupling of electromagnetic strength and a scalar mass of 1.1 TeV,  
a value chosen so as to produce sizeable shifts in $<P_2>$. This toy model 
yields the results shown in Table 5 where reasonable shifts in all the 
$<P_n>$ are apparent. In comparison to graviton KK graviton exchange, several 
differences with the scalar exchange case are immediately obvious: ($i$) the 
shifts in both $<P_1>$ and $<P_4>$ are larger in the scalar case whereas 
those for $<P_3>$ are significantly smaller by almost an order of magnitude. 
($ii$) While both $<P_{1,3}>$ are shifted, their {\it ratio} remains at its 
SM value 
for all flavors unlike in the case of gravity. It is clear that even this 
ad hoc toy model will not be confused with graviton KK tower exchange. 

\begin{table}
\centering
\begin{tabular}{|c|c|c|c|c|} \hline\hline
 $f$  &$<P_2>(10^{-2})$&$<P_4>(10^{-3})$&$<P_1>$  &$<P_3>(10^{-3})$  
\\ \hline \hline
$\mu,\tau$ & 7.03 & -7.55    & 0.236    & -2.93      \\
 $b$       & 4.23 & -6.41    & 0.188    & -2.34      \\
 $c$       & 5.78 & -7.04    & 0.248    & -3.08     \\
 $t$       & 2.35 & -5.87    & 0.148    & -1.83    \\ \hline\hline
\end{tabular}
\caption{Same as the previous table but now assuming the $s-$channel exchange 
of a 1.1 TeV scalar with 
universal couplings to all fermions as described in the text.}
\end{table}

\section{$e^+e^- \to W^+W^-$}

Can we use other processes to uniquely isolate the effects of graviton KK 
tower exchange? The other SM processes with large tree-level cross 
sections in which gravitons can be exchanged are $e^+e^- \to e^+e^-, 
\gamma\gamma, ZZ$, and $W^+W^-$ all of which involve $t-$ and/or 
$u-$channel exchanges. This would apparently disqualify them from further 
consideration. The $W^+W^-$ case is, however, special{\cite {gravww}} 
because ideally 
the offending $t-$channel $\nu$ exchange can be removed through the use of 
right-handed beam polarization leaving only the 
$s-$channel $\gamma,Z$ exchanges. 
The remaining purely right-handed 
SM cross section is only quadratic in $z$ and this will not 
change if, \eg, $Z'$ or new $s-$channel scalar contributions are also present. 
As we will discuss in detail below the difficulty in this case is 
that the left-handed cross section is much larger than 
the right-handed one so that the possibility of `contamination' from the wrong 
polarization state is difficult to eradicate unless very good control over the 
beam polarization is maintained. Apart from the problem of isolating the 
purely right-handed part of the cross section  
we might again conclude that the non-zero $<P_{3,4}>$ moments will be be a 
unique signature of graviton exchange for this process. Futhermore, since the 
pure gauge sector of the SM individually conserves $C$ and $P$, there are no 
terms in the cross section linear in $z$ and thus $<P_1>$ is expected to be 
zero in the SM and in many other NP extensions. Such terms are, however, 
generated by KK graviton exchange so that a non-zero value of $<P_1>$ is also a 
potential gravity probe.

Apart from new particle exchanges there is another source of NP that 
can modify the right-handed $W$-pair cross-section in a manner 
similar to gravity and would most likely be observed in lowest order as a 
dimension-8 operator (as is graviton KK tower exchange): anomalous gauge 
couplings(AGC){\cite {ac}}. As is well known AGC can be $C$ and/or $P$ 
violating; one that violates both $C$ and $P$ but is $CP$ conserving can (and 
does) produce non-zero $<P_{1,3,4}>$ moments. Decomposing the $WWV$ 
($V=\gamma,Z$) vertex in the most general way allowed by electromagnetic 
gauge invariance yields 7 different anomalous couplings for each $V$ with the 
corresponding  
form factors denoted by $f_i^V$. (When weighted by the sum over 
the $\gamma$ and $Z$ propagators in $e^+e^-\to W^+W^-$ 
these form factors are sometimes written as $F_i$, only two of 
which, $F_{1,3}$, are non-zero at the tree-level in the SM.) 
There is a single term in this general vertex expression with 
the correct $C$ and $P$ properties: that proportional to 
$f_5^V\epsilon^{\mu\alpha\beta\rho}(q^- -q^+)_\rho\epsilon_\alpha(W^-)
\epsilon_\beta(W^+)$ with $q^\pm$ the outgoing $W^\pm$ 
boson momenta, the $\epsilon$'s are their corresponding polarization 
vectors and $f_5^V$ 
being the relevant form factor. We note that even in the SM, though absent 
at the tree-level, this term 
is generated at one-loop from the usual fermion triangle anomaly 
graph. As we will see such a term will generate non-zero values for 
all of $<P_{1,3,4}>$. 

It appears that the possibility of non-zero AGC would 
contaminate our search for unique graviton exchange signatures. 
There is a way out of this dilemma; while gravity induces non-zero values 
for all of the $<P_{1,3,4}>$ from  
the angular distributions for $e^+e^-_R \to W^+W^-$ {\it independently}
of the final state $W$ polarizations, 
the $f_5^V \neq 0$ (\ie, $F_5$) couplings  
only contribute to the final state with mixed 
polarizations, \ie, transverse plus longitudinal, $TL+LT$. We recall that 
by measuring the angular distribution of the decaying $W$ relative to its 
direction of motion we can determine its state of polarization; here we do 
not differentiate the two possible states of transverse polarization. Writing 
\begin{equation}
{d\sigma_R\over {dz}} \sim \Sigma_{TT}+\Sigma_{LL}+\Sigma_{TL+LT}\,,
\end{equation} 
in the absence of $CP$ violation there are only 4 relevant $F_i$ and one finds 
\begin{eqnarray}
\Sigma_{TT} &\sim &  F_1^2(1-z^2) \nonumber \\
\Sigma_{LL} &\sim &\Big[F_3-(1-{2m_W^2\over {s}})F_1 +{\beta^2s\over {2m_W^2}}
F_2\Big]^2(1-z^2)\\
\Sigma_{TL+LT} &\sim &(F_3+\beta z F_5)^2(1+z^2)+2F_5(F_3+\beta z F_5)z(1-z^2)+
F_5^2(1-z^2)^2\,, \nonumber
\end{eqnarray}
where $\beta$ is the speed of the outgoing $W$'s. Here we see that in the 
SM both 
the $TT$ and $LL$ terms are proportional to $1-z^2$ while the $TL+LT$ term is 
proportional to $1+z^2$; no terms linear in $z$ are present. A non-zero 
$F_5$ induces additional terms in the case of the 
$TL+LT$ final state which now contains linear, cubic and quartic powers of 
$z$ similar to that generated by 
gravity. However, the $TT$ and $LL$ final states 
receive no such contributions. Thus 
observing non-zero values of $<P_{1,3,4}>$ (again, above backgrounds) for $W$ 
pairs in the $TT+LL$ final states produced by right-handed electrons 
{\it is} a signal for KK graviton tower exchange. Numerically the $TT$ 
fraction will dominate in the energy region of interest to us below. 

\begin{table}
\centering
\begin{tabular}{|c|c|c|c|c|} \hline\hline
     &$<P_2>(10^{-1})$&$<P_4>(10^{-3})$&$<P_1>(10^{-2})$&$<P_3>(10^{-2})$  
\\ \hline \hline
 SM                      & -2.14 & -6.94     & 0.0       & 0.0     \\
 $\Lambda_H=2$ TeV       & -2.16 & -8.88     & -4.88     & 2.46    \\
 SM'                     &  1.05 & -113.30   & 44.24     & -13.07    \\
 SM''                    & -1.11 & -41.33    & 14.30     & -4.22    \\
\hline\hline
\end{tabular}
\caption{Moments of the normalized $W$ pair production cross section assuming 
purely right-handed electrons and isolating the $TT+LL$ final states at 
$\sqrt s$=500 GeV. 
An angular cut $|\cos \theta|\leq 0.9$ has been applied. The SM prediction is 
compared with that for KK graviton tower exchange. Also shown is the SM 
prediction, labelled by SM'(SM'') for the case of $80\%$ right-handed 
$e^-$ and $60\%$ left-handed $e^+$ polarization(both 
beams with $90\%$ polarization.)}
\end{table}

To get an idea of the size of the graviton contributions to the $TT+LL$ part 
of the right-handed cross section, we show in Table 6 a comparison of the 
$<P_n>$ obtained in the SM and in the case with graviton KK tower exchange 
assuming $\Lambda_H=2$ TeV and $\sqrt s$=500 GeV. 
For later purposes, a cut of $|\cos \theta|\leq 0.9$ has been 
applied and an efficiency of $40\%$ has been assumed to both reconstruct 
the two $W$'s and to isolate the $TT+LL$ final state through their decay 
angular distribution. ISR has been ignored in these results. 
Here we see that the shifts for $<P_{1,3}>$ are quite 
large while those for $<P_2>$ are somewhat smaller as would be expected from 
the interference of the SM and KK gravity contributions. Provided purely 
right-handed electron beams were available, we can easily determine 
the graviton KK 
tower ID limit in this case. (Recall that we can now make use of all three of 
the moments $<P_{1,3,4}>$ in the $\chi^2$.) The results can be found in Fig.2. 
Note that the identification reach found in this extremely idealized 
situation of $100\%$ right-handed 
beam polarization is somewhat inferior to that found in 
the case of the $f \bar f$ final state, \ie, $\sim (4-5)\sqrt s$ at best. In 
any more realistic situation, especially when systematic effects are 
included, we expect this ID reach to significantly degrade.

\begin{figure}[htbp]
\centerline{
\includegraphics[width=9cm,angle=90]{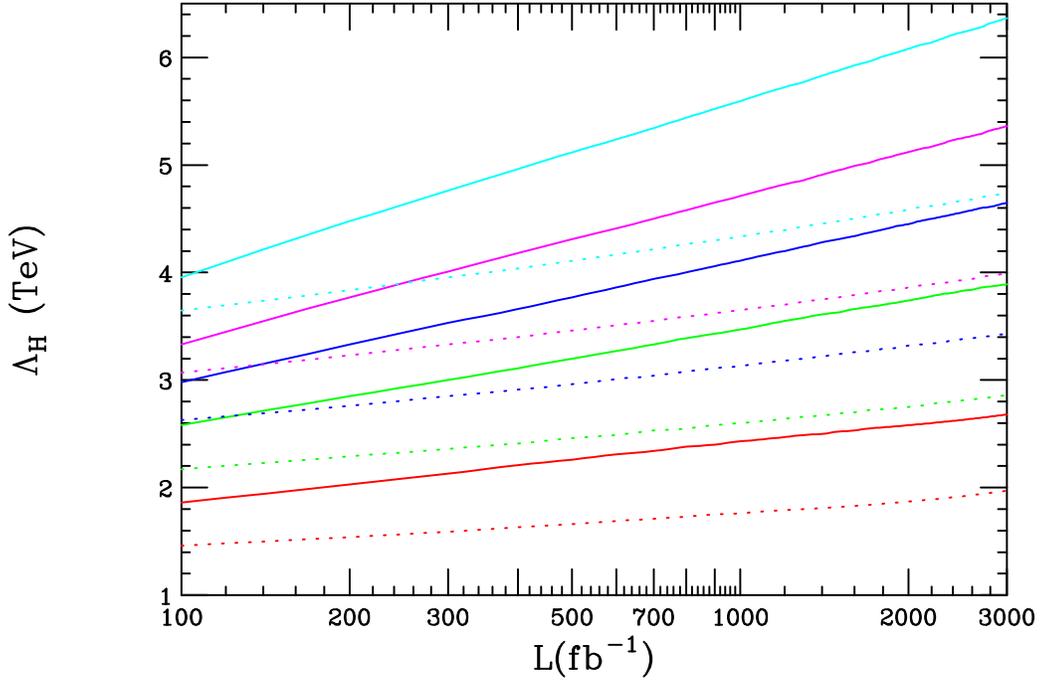}}
\vspace*{0.1cm}
\caption{Identification reach in $\Lambda_H$ as a function of the integrated 
luminosity for the process $e^+e^-_R\to W^+W^-$(solid), using only the $TT+LL$ 
final states. 
From bottom to top the curves are for $\sqrt s=0.5, 0.8,1, 1.2$ and $1.5$ TeV, 
respectively. The corresponding dotted curves are for the case of $80\%$ 
$e^-$ and $60\%$ $e^+$ polarization.}
\end{figure}

Perhaps, 
the closest we may be able to come to this idealized case experimentally 
it to assume double 
beam polarization taking the initial $e^-(e^+)$ as right(left)-handed as 
possible. Assuming polarizations of $80\%$ and $60\%$ for the $e^-$ and $e^+$ 
beams, respectively, increases the right-handed part of the cross section by 
a factor of 2.88 while reducing the left-handed part by a factor of 0.08. 
Unfortunately, the left-handed piece containing the $t$-channel $\nu$ 
exchange diagram is very large even after the cut of $|\cos \theta|\leq 0.9$ 
has been applied; recall that the left-handed part of the cross section has a  
large forward peak due to the $t-$channel exchange. 
Table 6 shows that with this degree of polarization (and 
even if both polarizations are unrealistically greater) the amount of 
contamination from the left-handed cross section is so large that making any 
claim to a unique identification of graviton exchange would be quite tenuous. 
We can still ask the same question as above: at what value of $\Lambda_H$ do 
the shifts in the $<P_n>$ correspond to a $5\sigma$ deviation from the SM? 
The answer shown in Fig.2, however, is no longer a unique 
identification of gravity but only a signal for the 
clear discovery of NP. Once contaminated by the left-handed cross section, 
other sources of NP, such as a $Z'$, can also lead to identical changes in the 
values of the moments.

The last possibility for salvaging this situation 
is to measure the $W^+W^-$ cross section with two or 
more sets of different beam polarizations and then attempting to extract the 
purely right-handed piece from these measurements, again keeping only the 
$TT+LL$ contributions. In the case of two 
polarized beams this is perhaps best demonstrated by examining what happens 
when we combine two sets of data: one with $P(e^-)=-80\%$ and $P(e^+)=60\%$ 
and the other with both polarizations 
flipped. (This is not necessarily the optimized choice of polarizations.) 
In comparison to the idealized purely right-handed case discussed 
above, here we suffer from having to be able to very precisely subtract the 
additional large backgrounds arising from 
the left-handed parts of the cross section. 
In addition, both reduced statistics (since the luminosity is 
divided between both measurements) and the systematic errors associated with 
the polarization uncertainties will lead to further reductions in the 
anticipated identification reach. In fact, assuming polarization uncertainties 
of $\delta P/P=0.3\%$ as above, we might expect these systematic effects to 
play an important role in the measurement error budget. 

Fig.3 shows the results of this analysis. Here we see that, as expected, the 
identification reach at large luminosities saturates due to the size of the 
systematic errors in extracting the right-handed piece of the cross section. 
The $5\sigma$ identification 
reach is found to be roughly $\sim 2.5\sqrt s$ for integrated luminosities 
of order $1~ab^{-1}$ which is far below that found for fermion pairs and the 
naive estimate we obtained in the case of purely right-handed 
$e^-$ polarization. It is unlikely that a more judicious choice of beam 
polarizations could drastically 
increase this reach and make it competitive with that 
found for the $f\bar f$ final state. 

\begin{figure}[htbp]
\centerline{
\includegraphics[width=9cm,angle=90]{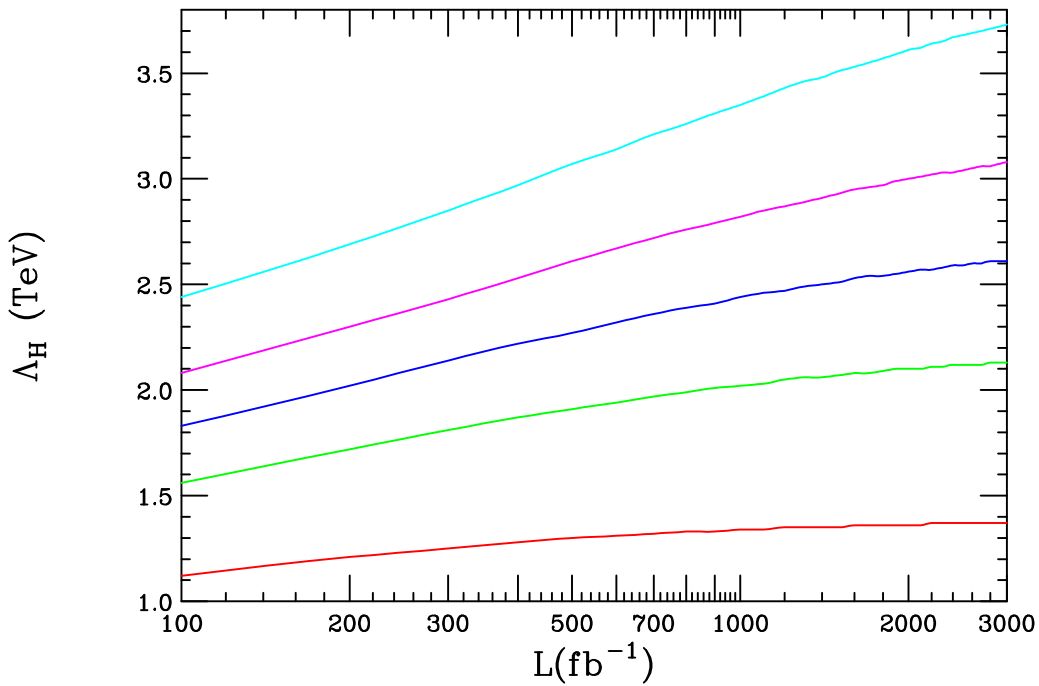}}
\vspace*{0.1cm}
\caption{Same as in the previous Figure but now using the combined analysis 
as discussed in the text.}
\end{figure}

\section{Summary and Conclusion}

Many new physics scenarios predict the existence of contact interaction-like 
deviations from SM cross sections at high energy $e^+e^-$ colliders. If  
such effects are observed our next task would be to identify their origin. 
In this paper we have suggested a technique by which the deviations induced 
by the exchange of KK gravitons can be uniquely isolated from all other 
possible sources by an examination of the angular moments of the polarization 
dependent cross sections 
employing Legendre polynomials. The technique 
is applicable when the `background' SM 
process proceeds only through $s-$channel exchanges or when it can be made 
to do so by a 
special choice of beam polarization(s). The canonical process to study is 
$e^+e^- \to f\bar f$ for $f\neq e$. In this case it was found that by 
combining several final states the  
graviton exchange contributions can be uniquely identified at the $5\sigma$ 
level for ADD mass scales as large as $(5-6)\sqrt s$ in the Hewett scheme. This 
compares rather favorably to the search reach of $(9-10)\sqrt s$ using this 
same process. The reaction $e^-_Re^+ \to W^+W^-$ for the $TT$ and $LL$ 
final states also proceeds only via 
$s-$channel exchange in the SM and can also be used to obtain a unique 
signature for graviton exchange. The $LT+TL$ modes, which we do not include, 
were shown to be capable of receiving graviton-like 
contributions from the $C$ and $P$ odd, $CP$ even anomalous trilinear 
couplings $f^V_5$. The difficulty in the case of the $W^+W^-$ final state 
is that $100\%$ polarized 
beams do not exist so that measurements made with different beam polarizations 
must be combined to extract the values of 
$d\sigma_R$ and as such, will suffer from sizeable systematic uncertainties. 
Though no attempt was made to 
optimize the choices of beam polarization, it was shown that even with both 
beams polarized the identification reach in this channel is somewhat below 
$\sim 2.5\sqrt s$ which is less than half that found 
for the $f\bar f$ final state. 
It is unlikely that optimization can lead to any sizeable improvement of this 
identification reach. 

Hopefully the effect of new contact interactions will be observed at the 
Linear Collider so that these new techniques can be employed.

\noindent{\Large\bf Acknowledgements}

The author would like to thank J.L. Hewett for discussion related to this work. 

%
\def\MPL #1 #2 #3 {Mod. Phys. Lett. {\bf#1},\ #2 (#3)}
\def\NPB #1 #2 #3 {Nucl. Phys. {\bf#1},\ #2 (#3)}
\def\PLB #1 #2 #3 {Phys. Lett. {\bf#1},\ #2 (#3)}
\def\PR #1 #2 #3 {Phys. Rep. {\bf#1},\ #2 (#3)}
\def\PRD #1 #2 #3 {Phys. Rev. {\bf#1},\ #2 (#3)}
\def\PRL #1 #2 #3 {Phys. Rev. Lett. {\bf#1},\ #2 (#3)}
\def\RMP #1 #2 #3 {Rev. Mod. Phys. {\bf#1},\ #2 (#3)}
\def\NIM #1 #2 #3 {Nuc. Inst. Meth. {\bf#1},\ #2 (#3)}
\def\ZPC #1 #2 #3 {Z. Phys. {\bf#1},\ #2 (#3)}
\def\EJPC #1 #2 #3 {E. Phys. J. {\bf#1},\ #2 (#3)}
\def\IJMP #1 #2 #3 {Int. J. Mod. Phys. {\bf#1},\ #2 (#3)}
\def\JHEP #1 #2 #3 {J. High En. Phys. {\bf#1},\ #2 (#3)}

\end{document}